\documentclass[a4paper,showpacs,preprintnumbers,nofootinbib,superscriptaddress,groupedaddress,11pt]{article}
\usepackage{graphicx}
\usepackage{dcolumn}
\usepackage{bm}
\usepackage{hyperref}
\usepackage[mathlines]{lineno}
\usepackage{amsfonts}
\usepackage{amsmath, amsfonts, amsthm, amssymb, amscd, mathrsfs}
\usepackage[mathcal]{euscript}

\newcounter{mnotecount}[section]

\renewcommand{\themnotecount}{\thesection.\arabic{mnotecount}}

\newcommand{\mnote}[1]
{\protect{\stepcounter{mnotecount}}$^{\mbox{\footnotesize
$
\bullet$\themnotecount}}$ \marginpar{
\raggedright\tiny\em
$\!\!\!\!\!\!\,\bullet$\themnotecount: #1} }

\hypersetup{colorlinks,urlcolor=blue}

\usepackage[portuguese,english]{babel}

\usepackage{xargs}
\usepackage{float}

\begin{document}

\title{Primordial magnetic fields in theories of gravity with non-minimal coupling between curvature and matter}

\author{Orfeu Bertolami$^{(1,*)}$, Maria Margarida Lima$^{(1,2)}$ and Filipe C. Mena$^{(3,4)}$ \\\\
{\small $^{(1)}$Departamento de F\'isica e Astronomia, Faculdade de Ci\^encias,}
\\
{\small Universidade do Porto, Rua do Campo Alegre 687, 4169-007, Porto, Portugal}
\\
{\small $^{(2)}$CEOS.PP, ISCAP, Polytechnic of Porto, Porto, Portugal}
\\
{\small $^{(3)}$Centro de An\'alise Matem\'atica, Geometria e Sistemas Din\^amicos,}
\\
{\small Instituto Superior T\'ecnico, Univ. Lisboa, Av. Rovisco Pais 1, 1049-001 Lisboa, Portugal}
\\
{\small $^{(4)}$Centro de Matem\'atica, Universidade do Minho, 4710-057 Braga, Portugal}
\\
{\small $^{(*)}$Corresponding author. Email: orfeu.bertolami@fc.up.pt}
}
	
%
%
%

	\date{\today}
	\maketitle
	\begin{abstract}
		The existence of magnetic fields in the universe is unmistakable. They are observed at all scales from stars to galaxy clusters. However, the origin of these fields remains enigmatic. It is believed that magnetic field seeds may have emerged from inflation, under certain conditions. This possibility is analised in the context of an alternative theory of gravity with non-minimal coupling between curvature and matter. We find, through the solution of the generalised Maxwell equations in the context of non-minimal models, that for general slow-roll inflationary scenarios with low reheating temperatures, $T_{RH}\simeq10^{10}$GeV, the generated magnetic fields can be made compatible with observations at large scales, $\lambda \sim 1 Mpc$.
	\end{abstract}
	
	
	\section{\label{Introduction}Introduction}

Magnetic fields play an important role in the dynamics of the astrophysical objects. For instance, the galactic magnetic field affects the dynamics of compact stars and, consequently, the star formation process \cite{Kronberg1994}. Furthermore, stars with extremely intense magnetic fields, $B\sim 10^{14}G$, are believed to be associated with recently observed rapid bursts \cite{Bochenek2020}. Although these fields are well understood, their cosmological origin remains an interesting open problem \cite{Barrow1997, Kari1998, Widrow2002, Massimo2004, Bamba2004, Giovannini2008}. 
	
Our galaxy, like many others, has a coherent magnetic field of $B\sim 10^{-6} G$ and a plausible explanation for these observed galactic magnetic fields is the dynamo effect \cite{Parker1979, Zeldivich1983, Ruzmainkin1988,Ferriere2009}. According to this mechanism, the differential rotation of galaxies exponentially enhances the magnetic field. However this is just an amplification mechanism and hence primordial magnetic fields must be considered. Considering that the dynamo effect operated during the entire life of the universe ($\sim$ 14 Gyears), then it is possible to amplify a seed field by a $e^{30}$ factor \cite{Turner1988}. Thus, this process allows to obtain the magnetic field strengths observed today from a seed field of approximately $B \sim 10^{-19} G$ \cite{Turner1988}.
	
Other lines of thought admit that the magnetic field in galaxies emerge due the compression of a primordial magnetic field during the collapse of the protogalactic cloud. In this case a stronger seed magnetic field is needed of at least $B \sim 10^{-9} G$ \cite{Turner1988}. For results about the magnetic field at intergalactic scales we refer the reader to Refs. \cite{Neronov2010, Tavecchio2010, Caprini2015}.

Primordial magnetogenesis thus remains an important area of research, aiming fundamentally to justify the large scale magnetic fields observed today (see Refs. \cite{Kandus2011,Durrer2013} for  excellent reviews). In fact, many studies have examined the evolution of magnetic fields in the radiation, dust and inflationary eras of the universe \cite{Kandus2011,Durrer2013}, as well as on the implications of magnetic fields in the evolution of cosmological gravitational waves \cite{Barrow2007, Axel2021} and possible imprints on the cosmic microwave background \cite{Kunze2013}.

Regarding the possible early universe scenarios for the creation of primordial magnetic fields, it is believed that, under specific conditions, the inflation era is the most promising set up. In fact the inflationary scenario of accelerated expansion \cite{Starobinsky1980, Guth1981, Linde1982, Albrecht1982} is still the current cosmological paradigm to solve the problems of the standard Big Bang cosmology, such as the horizon and flatness problems. The ability of this model is broader as, through the produced quantum fluctuations, inflation can ensure that the electromagnetic field is excited and the local magnetic flux is increased. Furthermore, using a mechanism similar to superadiabatic amplification, long wavelength modes ($\lambda \gtrsim H^{-1} $, where $H$ is Hubble expansion rate) are increased during the inflation and reheating eras.

However, during the inflationary period the magnetic field intensity decreases as $a^{-2}(t)$, where $a(t)$ is the scale factor of the flat Robertson-Walker metric. This is partly due to the conformal invariance of Maxwell's equations in General Relativity. Then, in order to create enduring magnetic fields during inflation in this context, the conformal symmetry of electromagnetism must be broken \cite{Turner1988, Garretson1992, Ratra1992, Dolgov1993,  Bertolami1999, Bertolami2005,Bamba2020,Chakraborty2022}.

This problem has been studied not only in the context of General Relativity but also e.g. in the Poincaré gauge theory \cite{Saketh2020}, supergravity and string theory \cite{Gasperini1995,Martin2008}, Gauss-Bonnet gravity \cite{Atmjeet2014} and torsion theories of gravity \cite{Saketh2020}. See also section 6 of \cite{Kandus2011} and references therein.

In this paper we will investigate whether an alternative theory of gravity with non-minimal coupling between curvature and matter, combined with inflation, can generate sizeable primordial magnetic fields. In this case, the breaking of the electromagnetic conformal invariance required for inflationary magnetogenesis arises naturally. In order to keep our calculations as general as possible we do not specify the scalar field potential driving inflation. We assume the slow-roll approximation though and the usual reheating phase at the end of inflation. We then calculate the ratio $r$ of the energy density in the magnetic field over the energy density in the background radiation fluid and compare our results with the known $r$-estimates needed to comply with the observational data. 
 Interestingly, we find that the generated magnetic fields in this theory can be made compatible with the observations.

This paper is organized in the following manner: In Section \ref{Chapter_model} we recall the main features of the theories of non-minimal coupling between curvature and matter. In Section \ref{Chapter_magnetic} we derive the generalised Maxwell equations in the context of the non-minimal coupling theories and analyze the evolution of the primordial magnetic field during inflation. Finally, we summarize our results in Section \ref{Chapter_conclusion}.
\section{\label{Chapter_model} The non-minimal coupling model}
	
Despite its mathematical beauty, several issues suggest that General Relativity (GR) is not the final theory of gravity \cite{Capoziello2011,Clifton2012, Bertol2011}. 
Well motivated alternatives to GR include higher-order curvature terms, such as $f(R)$ theories, as well as theories where matter and curvature are coupled non-minimally, as discussed next.	
\subsection{Non-minimal matter-curvature coupling theories}
	
GR rests on a principle of minimal coupling between curvature and matter which implies the covariant conservation of the energy-momentum tensor. However, it is possible to generalize the coupling providing an extra force in the geodesic equation that can mimic the dark matter effect on galaxies \cite{Bertolami2007}.
	
The action functional for those non-minimal coupling models is \cite{Bertolami2007}
\begin{equation}
	\label{action}
	S=\int \sqrt{-g}\bigg(\frac{1}{16\pi G}f_1(R)+ f_2(R) \mathcal{L}\bigg) d^4\mathbf{x},
\end{equation} 
where $f_1(R)$ and $f_2(R)$ are sufficiently smooth arbitrary functions of the Ricci curvature scalar $R$, whereas $g$ is the determinant of the 4-dimensional spacetime metric $g_{\mu\nu}$ and $\mathcal{L}$ a Lagrangian.
Varying the action with respect to the metric, we obtain the field equations: 
\begin{eqnarray}
	\label{field_equations}
	 \big( F_1(R)+16\pi G F_2(R) \mathcal{L} \big) G_{\mu\nu} &=& 8\pi G f_2(R)T_{\mu\nu}
	+\Delta_{\mu\nu}  \big( F_1(R)+16\pi G F_2(R) \mathcal{L} \big) \nonumber  \\ 
	&+&\frac{1}{2}g_{\mu\nu}\bigg( f_1(R)
	- \big(F_1(R)+16\pi G F_2(R) \mathcal{L}\big) R \bigg),
\end{eqnarray} 
where $F_i(R)=\frac{df_i(R)}{dR}$, $i\in \{1,2\}$, $\Delta_{\mu\nu}=\nabla_\mu \nabla_\nu - g_{\mu\nu}\Box$, $\Box=\nabla_\mu \nabla^\mu$ and $T_{\mu\nu}=\frac{-2}{\sqrt{-g}} \frac{\delta (\sqrt{-g} \mathcal{L}) }{\delta g^{\mu\nu}}$. GR is recovered for $f_1(R)=R$ and $f_2(R)=1$ as Eq. (\ref{field_equations}) reduces then to Einstein's field equations.
	
The trace of Eq. (\ref{field_equations}) is given by
\begin{eqnarray}
\label{trace-eq}
\big( F_1(R)+ 16\pi G F_2(R) \mathcal{L} \big) R -2 f_1(R)=8\pi G f_2(R) T-3\Box \big( F_1(R)+16\pi G F_2(R) \mathcal{L} \big),
\end{eqnarray}
where $T=g^{\mu\nu}T_{\mu\nu}$. 
Applying the contracted covariant derivative in Eq. (\ref{field_equations}) and using the contracted Bianchi identities, we obtain the relation
\begin{equation}
	\nabla_\mu T^{\mu\nu}=\frac{F_2(R)}{f_2(R)}(\mathcal{L} g^{\mu\nu}-T^{\mu\nu})\nabla_\mu R. 
	\label{non_conservative_T}
\end{equation}
Thus, in general, the energy-momentum tensor is not covariantly conserved. It is easily observed that in the GR limit the previous equation becomes the usual conservation equation.
From Eq. (\ref{non_conservative_T}), results that an extra force, $f^\mu$, appears in the geodesic equation e.g. of a perfect fluid \cite{Bertolami2007}: 
\begin{equation}
	f^\mu =\frac{1}{\rho+p} \bigg( \frac{F_2(R)}{f_2(R)}(\mathcal{L}-p)\nabla_{\nu}R +\nabla_{\nu}p \bigg)h^{\mu\nu}, 
\end{equation}
where $h^{\mu\nu}=g^{\mu\nu}-u^\mu u^\nu$ is the projection operator with respect to the $4$-velocity $\mathbf {u}$ of the fluid with energy density $\rho$ and pressure $p$. 	
\subsection{Inflation in non-minimal coupling theories}
	
For the purpose of considering the generation of primordial magnetic fields during inflation in the context of the non-minimally coupled curvature-matter theories of gravity, we describe some general features of this mechanism (see also \cite{Gomes2017}). 

Inflation is driven by a scalar field, the inflaton, whose Lagrangian density is given by: 
\begin{equation}
		\mathcal{L}_\phi = -\frac{1}{2} \partial_\mu \phi \partial^\mu \phi-V(\phi),
	\label{lagrang_inflaton}
\end{equation}
where $V(\phi)$ is the potential of the inflaton.

We recall that during inflation the potential energy of the scalar field dominates the spacetime dynamics leading to an accelerated expansion. So, if there is a region of the universe where the field is homogeneous and rolls down its potential slowly, we get an accelerated expansion that dilutes any energy gradient of the scalar field. Regarding studies about pre-inflationary homogeneization see e.g. Ref. \cite{Alho2011} and references therein.

For the geometry we consider a spatially homogeneous and isotropic universe described by the flat Robertson-Walker metric:
\begin{equation}
	ds^2=-dt^2+a^2(t) \big( dx^2+dy^2+dz^2 \big). 
	\label{cartesian_metric}
\end{equation}
Using the definition of the energy-momentum tensor with the Lagrangian density (\ref{lagrang_inflaton}), we get
\begin{equation}
	T_{\mu\nu}=\partial_\mu \phi \partial_\nu \phi -g_{\mu\nu} \big( \frac{1}{2}\partial_\alpha \phi \partial^\alpha \phi+V(\phi) \big)
\end{equation}	
and making the usual identifications with the perfect fluid energy-momentum tensor, compatible with the metric \eqref{cartesian_metric}, one obtains the well-known expressions
\begin{eqnarray}
\label{density}
	\rho_\phi =\frac{1}{2}\dot{\phi}^2+ V(\phi),~~~~~
	p_\phi =\frac{1}{2}\dot{\phi}^2 - V(\phi)  
\end{eqnarray}
for the energy-density and pressure of the scalar field, where the dot denotes differentiation with respect to $t$.

In turn, the evolution equation of the inflaton in the context of non-minimal coupling theory is given by:
\begin{equation}
	\ddot{\phi}+3H\dot{\phi}+\frac{d V(\phi)}{d \phi}=-\frac{F_2(R)}{f_2(R)} \dot{R} \dot{\phi},
\end{equation}
where $H=\dot a/a$ is the Hubble function. So, comparing to GR, the non-minimal coupling between curvature and matter induces an additional friction term on the right-hand-side of the previous scalar field equation. 
	
We recall that the inflationary period will last as long as the kinetic energy of the scalar field is negligible, that is, during the phase where the potential is flat. When the potential becomes steeper and the field moves faster, inflation will end. After that, the inflaton potential energy will be transferred to radiation and ordinary matter.

Consider then fluid matter with density $\rho$ and pressure $p$. 	
If we consider the time component of the non-conservation equation (\ref{non_conservative_T}), it is possible to find the relation: 
\begin{equation}
	\dot{\rho} + 3H(\rho+p)=-\frac{F_2(R)}{f_2(R)} (\rho+p)\dot{R}.
\end{equation}
Similarly to the GR case, we can write a linear equation of state as $p=w\rho$, and in this case
\begin{equation}
	\frac{\dot{\rho}}{\rho}=-(1+w) \bigg( 3\frac{\dot{a}}{a}+\frac{F_2(R) \dot{R}}{f_2(R)} \bigg). 
\end{equation}

If $w=-1$, i.e. in the de Sitter era, the previous equation reduces to $\rho= \rho_0=const.$ as in GR.
	
If $w\neq-1$, we obtain 
\begin{equation}
	\rho=\rho_i \bigg( \frac{a_i}{a} \bigg)^{3(1+w)} \bigg( \frac{f_2(R_i)}{f_2(R)} \bigg)^{1+w}, 
\end{equation}
where the index $i$ denotes initial values.  
A sub-case of interest here is the case of a radiation fluid, $w=1/3$, which gives
\begin{equation}
	\rho_\gamma=\rho_i \bigg( \frac{a_i}{a} \bigg)^{4} \bigg( \frac{f_2(R_i)}{f_2(R)} \bigg)^{\frac{4}{3}}. 
	\label{density_radiation}
\end{equation}
The time-time component of Eq. (\ref{field_equations}) corresponds to the modified Friedmann equation that takes the form \cite{Gomes2017, Bertolami2014}: 
\begin{eqnarray}
	\label{Friedmann_eq_modif}
	H^2&=&\frac{1}{6  \big( F_1(R)+16\pi G F_2(R)\mathcal{L}_\phi \big)} \bigg( 16\pi G \rho_\phi f_2(R)
	-6 H \frac{\partial }{\partial t}\big( F_1(R)+16\pi G F_2(R)\mathcal{L}_\phi \big)\nonumber\\
	&-&f_1(R) +\big( F_1(R)+16\pi G F_2(R)\mathcal{L}_\phi \big)R \bigg). 
\end{eqnarray}	
In turn, we can write the modified Raychaudhuri equation, using Eq. (\ref{field_equations}), as
\begin{eqnarray}
	\label{Raychaudhuri_eq_modif}
        \big( F_1+16\pi G F_2\mathcal{L}_\phi \big) \big( 2\frac{\ddot{a}}{a}+H^2\big)&=&-8\pi G p_\phi f_2-
	 \frac{d^2}{d t^2} \big( F_1+16\pi G F_2\mathcal{L}_\phi \big)\nonumber \\
	 &-&3 H\frac{d}{dt} \big( F_1+16\pi G F_2\mathcal{L}_\phi \big) 
	- \frac{1}{2}\bigg( f_1 - \big( F_1+16\pi G F_2\mathcal{L}_\phi \big)R \bigg).\nonumber
\end{eqnarray}
In this scenario, in order for inflation to take place, the inflaton field must mimic a cosmological constant. In that case, the kinetic energy is negligible when compared to the potential energy and 
one has $p_\phi \simeq -\rho_\phi$. Then, using Eqs. (\ref{density}), the well-known slow-roll conditions are required:
\begin{equation}
	\frac{1}{2}\dot{\phi}^2 \ll V(\phi)~~~\text{and}~~~\ddot{\phi} \ll 3H\dot{\phi}
\end{equation}
and ensure that the inflaton's motion is sufficiently damped to allow for the accelerated expansion of the universe.
	
Taking into account the above conditions, then the time-time and space-space components of Eq. (\ref{field_equations}) can be written respectively as:
\begin{equation}
	f_2(R) \rho_\phi=3(FH^2+H\dot{F})+\frac{1}{2} \big(8\pi G f_1(R)-RF\big)
\end{equation}
and 
\begin{equation}
	f_2(R) p_\phi=-3(FH^2+H\dot{F})-\frac{1}{2} \big(8\pi G f_1(R)-RF\big)-2F\dot{H}-\ddot{F}
\end{equation}
where $F=\big(8\pi G F_1(R)-2F_2(R)\rho_\phi\big)$.
From these relations and the slow-roll approximation, it follows that the time derivatives of the curvature scalar $R$ and the matter density $\rho$ vanish.
	
Similarly to \cite{Gomes2017} let us consider that 
\begin{equation}
	f_1(R)=R.
\end{equation}
In this way we are isolating the effect of the non-minimal coupling between matter and curvature in order to understand its impact.
Thus, it is possible to simplify the modified Friedmann equation (\ref{Friedmann_eq_modif}) in the slow-roll regime to get
\begin{equation}
	H^2= \bigg(\frac{8\pi G f_2(R)}{1+16\pi G\rho_\phi F_2(R)}\bigg) \frac{\rho_\phi}{3}.
	\label{Friedmann_eq_modif_2}
\end{equation}
So, in this case, the modified Friedmann equation depends not only on the energy density of the inflaton (including the potential $V(\phi)$) but also on the non-minimal coupling function $f_2(R)$ considered.
	
\subsection{Cubic model}
	
On quite general grounds, the  non-minimal coupling can be written as a linear combination of powers of the scalar curvature, that is
\begin{equation}
	f_2(R)=\sum_{n=-\infty}^{+\infty}a_n \bigg( \frac{R}{\overline{R}_{n}} \bigg)^n=\sum_{n=-\infty}^{+\infty} \varUpsilon_n R^n, 
\end{equation}
where $a_0=\varUpsilon_0=1$ and $a_n$ and $\overline{R}_{n}$ are constants characteristic of the specific phase of the universe under consideration, and $\varUpsilon_n$ can be seen as a coupling constant that also depends on the phase.
	
As discussed in Ref. \cite{Gomes2017}, only powers of the curvature greater than quadratic lead to a non-trivial behaviour of the modified Friedmann equation. Thus, we choose: 
\begin{equation}
	f_2(R)=1+\xi R^3, 
	\label{non_min_func}
\end{equation}
where $\xi \geq 0$ parameterizes the deviation from GR. Given that 
\begin{equation}
	R=6\bigg( \frac{\ddot{a}}{a}+H^2 \bigg),
\end{equation}
considering the slow-roll regime and using \eqref{Friedmann_eq_modif_2} together with Eq. (\ref{non_min_func}), we get 
\begin{equation}
	R\simeq 12H^2.
	\label{R}
\end{equation}
Replacing the previous result and the non-minimal coupling function (\ref{non_min_func}) in the modified Friedmann equation (\ref{Friedmann_eq_modif_2}) we obtain:
\begin{equation}
	H^2(\xi,\rho)=\frac{1}{12}\frac{\bigg( 36\xi^2(8\pi G \rho)^3+\sqrt{6\xi^3(8\pi G \rho)^3\big( 216\xi (8\pi G \rho)^3 +1\big)} \bigg)^\frac{2}{3}-8\pi G  6^\frac{1}{3}\xi \rho}{8\pi G 6^\frac{2}{3}\xi \rho \bigg( 36\xi^2(8\pi G \rho)^3+\sqrt{6\xi^3(8\pi G \rho)^3\big( 216\xi (8\pi G \rho)^3 +1\big)} \bigg)^\frac{1}{3}}.
	\label{Friedmann_eq_modified}
\end{equation}
If we consider an expansion around $\xi=0$, thus accessing the deviation from the usual Friedmann equation, then
\begin{equation}
	H^2(\xi,\rho)= 8\pi G \frac{\rho}{3} - 32 (8\pi G)^4 \frac{\rho^4}{3}\xi +\mathcal{O}(\xi^2),
\end{equation}
where it is easy to check the GR limit $H^2(0,\rho)=8\pi G \rho/3$.
		
In the de Sitter phase, it is possible to exactly solve the modified Friedmann equation (\ref{Friedmann_eq_modified}), since the density $\rho=\rho_0$  is constant. So,  the scale factor takes the form: 
\begin{equation}
	a(t)= a_ie^{f(\xi)(t-t_0)},
\end{equation}
where $f(\xi)=H(\xi,\rho_0)$.
If we expand the previous equation around $\xi=0$ we get
\begin{equation}
	f(\xi)=H_0-6^4H^7_0\xi+\mathcal{O}(\xi^2).
	\label{Taylor_f}
\end{equation} 
where $H_0:=H(0,\rho_0)$ corresponds to the GR value.
	
	
\section{\label{Chapter_magnetic}  Primordial Magnetic field}
  
This section contains our main results. We first derive the generalised Maxwell equations, in the context of our non-minimal coupling model, and obtain an explicit solution for the evolution of the magnetic field on a flat Robertson-Walker background. Then we estimate the variation of the magnetic field intensity during the inflationary era and confront our results with observational data.

\subsection{Maxwell equations in the non-minimal coupling model}

We will use the metric \eqref{cartesian_metric} written in conformal time $\tau$ as: 
\begin{equation}
	ds^2=a^2(\tau)\big( -d\tau^2+dx^2+dy^2+dz^2 \big),
	\label{conformal_metric}
\end{equation}
where $dt=ad\tau$.
	
Recall that the electromagnetic Lagrangian density in the absence of a current is
\begin{equation}
	\mathcal{L}=-\frac{1}{4} F_{\mu\nu}F^{\mu\nu}, 
	\label{eletromagnetic_lagrangian}
\end{equation}
where $F^{\mu\nu}$ is the Faraday tensor, and that for the above metric
\begin{equation}
	F_{\mu\nu}= a^2	\begin{bmatrix}
		0&-E_x&-E_y&-E_z\\
		E_x&0&B_z&-B_y\\
		E_y&-B_z&0&B_x\\
		E_z&B_y&-B_x&0
	\end{bmatrix},
\end{equation}
where $E_j$ and $B_j$ denote the components of the electric and magnetic fields.
	
The energy-momentum tensor of the electromagnetic field is given by:
\begin{equation}
	T_{\mu\nu}=F_{\mu\alpha}F_\nu^\alpha-\frac{1}{4}g_{\mu\nu} F_{\alpha\beta}F^{\alpha\beta}.
\end{equation}
In order to obtain the Maxwell equations, we consider the action (\ref{action}) with the Lagrangian density (\ref{eletromagnetic_lagrangian}) and vary with respect to the 4-potencial $A_\mu=(\Phi, \mathbf{A})$, where $\Phi$ is the electric potential and $\mathbf{A}$ the vector potential.  

So we obtain the so-called inhomogeneous Maxwell equations
\begin{equation}
	\nabla_\mu \big(f_2(R) F^{\mu\nu} \big)=0,
	\label{Maxwell_eq_1_1}
\end{equation}
which, explicitly, give the following four equations:
\begin{eqnarray*}
	\frac{\partial}{\partial x} E_x+\frac{\partial}{\partial y} E_y+\frac{\partial}{\partial z} E_z&=&0,\\
	\frac{1}{a^2 f_2(R)}\frac{\partial}{\partial \tau} \big(a^2 f_2(R) E_x\big)+\frac{\partial}{\partial z} B_y-\frac{\partial}{\partial y} B_z&=&0,\\
	\frac{1}{a^2 f_2(R)} \frac{\partial}{\partial \tau}  \big(a^2 f_2(R) E_y\big)+\frac{\partial}{\partial x} B_z-\frac{\partial}{\partial z} B_x&=&0,\\
	\frac{1}{a^2 f_2(R)} \frac{\partial}{\partial \tau} \big(a^2 f_2(R) E_z\big)+\frac{\partial}{\partial y} B_x-\frac{\partial}{\partial x} B_y&=&0. 
\end{eqnarray*}
In turn, the homogeneous Maxwell equations are
\begin{equation}
	\nabla_\mu \tilde F^{\mu\nu}=0,
	\label{Maxwell_eq_2}
\end{equation}
where $\tilde F^{\mu\nu}=\frac{1}{2}F_{\alpha\beta}\epsilon^{\alpha \beta \mu \nu}$ and $\epsilon_{\alpha \beta \mu \nu}$ is the Levi-Civita tensor. The equations \eqref{Maxwell_eq_2} can written explicitly as
\begin{eqnarray*}
	\frac{\partial}{\partial x} B_x+\frac{\partial}{\partial y}  B_y+\frac{\partial}{\partial z} B_z&=&0,\\
	\frac{1}{a^2 } \frac{\partial}{\partial \tau} \big(a^2 B_x\big)+\frac{\partial}{\partial y} E_z-\frac{\partial}{\partial z} E_y&=&0,\\
	\frac{1}{a^2} \frac{\partial}{\partial \tau} \big(a^2 B_y\big)+\frac{\partial}{\partial z} E_x-\frac{\partial}{\partial x} E_z&=&0,\\
	\frac{1}{a^2 } \frac{\partial}{\partial \tau} \big(a^2 B_z\big)+\frac{\partial}{\partial x} E_y-\frac{\partial}{\partial y} E_x&=&0. 
\end{eqnarray*}
So we can gather the Maxwell equations as $\nabla \cdot \mathbf{B}=0$ and $\nabla \cdot \mathbf{E}=0$ together with
\begin{eqnarray}
	\frac{1}{a^2 f_2(R)} \big(a^2 f_2(R) \mathbf{E}\big)^\prime-\nabla\times \mathbf{B}&=&0,\\
	\label{Max_Eq_2}
	\frac{1}{a^2 } \big(a^2 \mathbf{B}\big)^\prime+\nabla\times \mathbf{E}&=&0, 
	\label{Max_Eq_1}
\end{eqnarray}
where the prime denotes differentiation with respect to conformal time, and the divergence and rotational operators are the usual ones in euclidean space. 
It is easy to verify that in the minimal coupling regime, i.e. for $f_2(R)=1$, we recover the usual Maxwell equations in GR (see e.g. \cite{Bertolami1999}).
	
Applying the curl to Eq. (\ref{Max_Eq_2}) we get
\begin{equation}
	\frac{1}{a^2 f_2(R)}\big(a^2 f_2(R) \nabla\times\mathbf{E}\big)^\prime-\nabla(\nabla \cdot \mathbf{B})+\nabla^2\mathbf{B}=0
\end{equation}
and using the remaining equations we find
\begin{equation}
\label{eq3}
(a^2 f_2 \mathbf{B})^{\prime\prime}+\left(\left(\ln{f_2}\right)^\prime a^2 f_ 2 \mathbf{B}\right)^\prime -\nabla^2 (a^2 f_2 \mathbf{B})=0.
\end{equation}
In general, this equation coupled to the modified Friedman equation can be solved numerically. Nevertheless, in order to get an estimate of the solution during  inflation we can neglect the middle term (since $(\ln{f_2})^\prime\approx 0$) and apply of the Fourier transform.
So, defining 
\begin{equation}
	\mathbf{F}_k(\tau)=a^2 f_2(R) \int e^{-\mathbf{k}\cdot\mathbf{x}}\mathbf{B} d{\mathbf x}
\end{equation}
we obtain
\begin{equation}
	\mathbf{F}^{\prime\prime}_k(\tau) + k^2 \mathbf{F}_k(\tau)=0, 
\end{equation} 
where $\mathbf {F} _k$ can be seen as a measure of the magnetic flux associated with the comoving scale $\lambda \sim k^{-1}$. 	
We can then write the magnetic field as
\begin{equation}
	\mathbf{B}=\frac{1}{a^2f_2(R)}\int \mathbf {F} _k(\tau) e^{i{\mathbf k}\cdot{\mathbf x}}d{\mathbf k}. 
\end{equation}
and since the integral is bounded, we get
\begin{equation}
	B \propto \frac{1}{a^2 f_2(R)}, 
	\label{Magnetic}
\end{equation}
where $B$ is the intensity of the magnetic field. 
		
\subsection{Magnetic field during inflation}

In this subsection we start by estimating the ratio between the magnetic field intensity at the beginning at the end of inflation when reheating occurs. Then we consider the ratio $r$ between the magnetic field density and the background radiation density and  estimate its value $r_{RH}$ at the reheating phase. After that, we use the Turner-Widow relation between $r_{RH}$ and $r$ to estimate the latter. Finally, we compare our estimates with the known $r$ values that are needed in order to comply with the observations. In order to do that we rely on estimates about the amplification of a seed magnetic field resulting from inflation until the present, from the dynamo effect and the compression effect.

We note that we do not compute the specific value for the necessary initial seed magnetic field at the beggining of inflation. Instead we focus on the effect that inflation might have on this seed field, in particular on whether the dillution effect is mild enough in order to comply with the present data.
	
Let us now focus on the result of Eq. (\ref{Magnetic}). The dilution factor of the magnetic field during inflation is given by
\begin{equation}
	\frac{B_e}{B_i}\simeq \bigg( \frac{a_i}{a_e} \bigg)^2\frac{f_2(R_i)}{f_2(R_e)},
\end{equation}
where the subindices $i$ and $e$ denote 	initial and final values, and as $a_e \sim e^{60} a_i$ we have
\begin{equation}
	\frac{B_e}{B_i}\simeq 10^{-53}\frac{f_2(R_i)}{f_2(R_e)}.
	\label{factor_decay}
\end{equation}
Let us assume that $f_2(R)$ is dominated by the cubic curvature term at the time of reheating (RH) that is: 
\begin{equation}
	f_{2_{RH}}(R)\simeq\varsigma R^3,
\end{equation}
where $\varsigma$ is a constant. 
	
We consider now that at the end of inflation the typical energy scale is the one related to the necessary reheating of the universe, 
\begin{equation}
H_{RH} \simeq \frac{\pi}{\sqrt{90}} \frac{T_{_{RH}}^2}{M},
\end{equation}
where $T_{_{RH}}$ is the reheating temperature and $M=(\sqrt{8\pi G})^{-1}$ is the reduced Planck mass.	
So, by Eq. (\ref{R}), 
\begin{equation}
	R_{RH} \simeq \frac{2 \pi^2}{15} \frac{T_{_{RH}}^4}{M^2}.
	\label{R_RH}
\end{equation}
Therefore, we can write
\begin{equation}
	\frac{B_{RH}}{B_i} \simeq 10^{-53} \bigg(\frac{H_I}{H_{RH}}\bigg)^6,
\end{equation}
where $H_I\simeq \frac{ \Delta^2}{M}$ is the expansion rate at inflation and we assume typically that $\Delta\simeq 10^{-3} M$ so that
\begin{equation}
	\frac{B_{RH}}{B_i} \simeq 10^{-53} \bigg(\frac{\Delta}{T_{RH}}\bigg)^{12}.
\end{equation}
If we assume that the reheating temperature is between the typical values $10^{-8}M \lesssim T_{RH} \lesssim 10^{-4}M$ (see e.g. Ref. \cite{Bertolami1999}), we can estimate that
\begin{equation}
	10^{-41}\lesssim\frac{B_{RH}}{B_i}\lesssim10^{7},
	\label{ratio}
\end{equation}
for magnetic fields coherent over $1 Mpc$ which is the typical length scale due to inflation. 

It is usual to consider the quantity
\begin{equation}
r=\frac{\rho_B}{\rho_\gamma},
\end{equation}
for an estimate of the resulting magnetic field, 
where $\rho_B=\frac{B^2}{8\pi}$ and $\rho_\gamma=\frac{\pi^2}{15}T^4$.

At present, for galaxies, $r=1$, hence $r$ as low as $10^{-34}$ can account for the observed galactic fields if the dynamo amplification mechanism is invoked \cite{Kandus2011}. On the hand, with the compression mechanism $r$ can be as low as $10^{-8}$ to be compatible with observations \cite{Turner1988}.

Considering now $\rho_{\gamma_i}=\frac{\pi^2}{15}T_i^4$ and that $r_i=10^\chi$, where $\chi$ is an arbitrary power, then from Eq. (\ref{ratio}) we get
\begin{equation}
	10^{-82+\chi}\lesssim r_{_{RH}} \lesssim 10^{14+\chi}.
	\label{ratio2}
\end{equation}
To obtain $r$ from $r_{_{RH}}$ several assumptions must be considered with respect the conductivity of the early universe. A partial analysis can be found in \cite{Turner1988,Bertolami1999} in terms of the temperature $T_{*}=\min\{ (T_{RH}\Delta)^{1/2}, (T_{RH}^2M)^{1/3} \}$, below which the universe becomes a good conductor: 
\begin{equation}
	r\approx\bigg( \frac{T_{*}}{M} \bigg)^{-8/3}r_{_{RH}}. 
	\label{eq_r}
\end{equation}
Therefore, combining Eqs. (\ref{ratio2}) and (\ref{eq_r}), we get: 
\begin{equation}
	10^{-68+\chi}\lesssim r \lesssim 10^{28+\chi},
\end{equation}
from which we conclude that $\chi$ can be as low as $-62$ if the amplification of primordial magnetic fields is performed through the dynamo effect. If, on the other hand, the compression mechanism is invoked then $\chi \geq -36$. 

\section{Conclusion}

\label{Chapter_conclusion} 

In the typical Big Bang models within General Relativity, primordial magnetic fields get dramatically diluted by the early universe inflationary expansion and cannot not account for the present observations. Alternative theories of gravity with a non-minimal coupling between curvature and matter can break the conformal invariance of Maxwell's equations and may provide a viable mechanism for a sizable magnetic field.

 We have then studied the variation of the magnetic field during inflation in the context of a model with non-minimal curvature-matter coupling. 
These theories include arbitrary functions of the Ricci curvature, $f_1(R)$ and $f_2(R)$ in the action. Keeping the closedness to GR and motivated by recent studies \cite{Gomes2017} we have considered $f_1(R)$ linear and $f_2(R)$ cubic in $R$. We have then derived the generalised Maxwell's equations in a flat Robertson-Walker metric and found that the magnetic field evolves as $B\propto 1/(a^2 f_2(R))$.

In the inflationary scenario we considered the slow-roll approximation and a reheating phase but we did not specify the scalar field potential, for the sake of generality. 
We have computed the ratio between the intensity of the magnetic field at the end and at the beginning of inflation.
For the typical inflation scales and reheating temperatures, we found that that ratio gives $10^{-41}\lesssim B_{RH}/B_i\lesssim10^{7}$ at large scales, $\lambda \sim 1 Mpc$.

We have then taken the initial density ratio $r=10^\chi$ between the energy densities associated with the magnetic field and the cosmological radiation fluid. Taking into account the present observations of the magnetic field in galaxies, we concluded that $\chi\ge -62$ if the amplification of primordial magnetic fields results from the dynamo effect and $\chi \geq -36$ if the compression mechanism is assumed. Our results suggest that inflation models with low reheating temperatures are required.

This study supports the idea that primordial magnetic fields generated within non-minimal coupling models can be made compatible with observations. 
\\

\noindent
{\bf Acknowledgements:} FCM thanks support from FCT/Portugal through CAMGSD, IST-ID, project UIDB/04459/2020 and UIDP/04459/2020 as well as CMAT, Univ. Minho, through project UIDB/00013/2020 and UIDP/00013/2020 and FEDER Funds COMPETE.
\\

\noindent
Data sharing not applicable to this article as no datasets were generated or analysed during the current study. 

\bibliographystyle{JHEP}
\bibliography{references}

\providecommand{\href}[2]{#2}\begingroup\raggedright\begin{thebibliography}{10}

\bibitem{Kronberg1994}
P.P.~Kronberg, \emph{Extragalatic magnetic fields}, {\emph{Rep. Prog. Phys.}
  {\bfseries 57} (1994) 325}.

\bibitem{Bochenek2020}
C.D.~Bochenek, V.~Ravi, K.V.~Belov, G.~Hallinan, J.~Kocz, S.R.~Kulkarni et~al.,
  \emph{A fast radio burst associated with a galactic magnetar}, {\emph{Nature}
  {\bfseries 587} (2020) 59}.

\bibitem{Barrow1997}
J.D.~Barrow, P.G.~Ferreira and J.~Silk, \emph{Constraints on a primordial
  magnetic field}, {\emph{Phys. Rev. Lett.} {\bfseries 78} (1997) 3610}.

\bibitem{Kari1998}
K.~Enqvist, \emph{Primordial magnetic fields}, {\emph{International Journal of
  Modern Physics D} {\bfseries 7} (1998) 331}.

\bibitem{Widrow2002}
L.M.~Widrow, \emph{Origin of galactic and extragalactic magnetic fields},
  {\emph{Rev. Mod. Phys.} {\bfseries 74} (2002) 775}.

\bibitem{Massimo2004}
M.~Giovannini, \emph{The magnetized universe}, {\emph{International Journal of
  Modern Physics D} {\bfseries 13} (2004) 391}.

\bibitem{Bamba2004}
K.~Bamba and J.~Yokoyama, \emph{Large-scale magnetic fields from dilaton
  inflation in noncommutative spacetime}, {\emph{Phys. Rev. D} {\bfseries 70}
  (2004) 083508}.

\bibitem{Giovannini2008}
M.~Giovannini, \emph{Magnetic fields, strings and cosmology}, {\emph{Lecture
  Notes in Physics} {\bfseries 737} (2006) 863}.

\bibitem{Parker1979}
E.N.~Parker, \emph{Cosmological Magnetic Fields. Their origin and their
  activity}, Oxford University Press (1979).

\bibitem{Zeldivich1983}
Y.B.~Zeldovich, A.A.~Ruzmaikin and D.D.~Sokoloff, \emph{Magnetic fields in
  astrophysics}, {\emph{Gordon and Breach, New York} (1983) }.

\bibitem{Ruzmainkin1988}
A.A.~Ruzmaikin, A.A.~Shukurov and D.D.~Sokoloff, \emph{Magnetic fields of
  galaxies}, {\emph{Kluwer, Dordrecht} (1988) }.

\bibitem{Ferriere2009}
K.~Ferri\`ere, \emph{Interstellar magnetic fields in the galactic center
  region}, {\emph{Astronomy \& Astrophysics} {\bfseries 505} (2009) 1183}.

\bibitem{Turner1988}
M.S.~Turner and L.M.~Widrow, \emph{Inflation-produced, large-scale magnetic
  fields}, {\emph{Phys. Rev. D} {\bfseries 37} (1988) 2743}.

\bibitem{Neronov2010}
A.~Neronov and I.~Vovk, \emph{Evidence for strong extragalactic magnetic fields
  from fermi observations of tev blazars}, {\emph{Science} {\bfseries 328}
  (2010) 73}.

\bibitem{Tavecchio2010}
F.~Tavecchio, G.~Ghisellini, L.~Foschini, G.~Bonnoli, G.~Ghirlanda and
  P.~Coppi, \emph{The intergalactic magnetic field constrained by fermi/large
  area telescope observations of the tev blazar 1es 0229+200}, {\emph{Monthly
  Notices of the Royal Astronomical Society: Letters} {\bfseries 406} (2010)
  L70}.

\bibitem{Caprini2015}
C.~Caprini and S.~Gabici, \emph{Gamma-ray observations of blazars and the
  intergalactic magnetic field spectrum}, {\emph{Phys. Rev. D} {\bfseries 91}
  (2015) 123514}.

\bibitem{Kandus2011}
A.~Kandus, K.E.~Kunze and C.G.~Tsagas, \emph{Primordial magnetogenesis},
  {\emph{Physics Reports} {\bfseries 505} (2011) 1}.

\bibitem{Durrer2013}
R.~Durrer and A.~Neronov, \emph{Cosmological magnetic fields: their generation,
  evolution and observation}, {\emph{The Astronomy and Astrophysics Review}
  {\bfseries 21} (2013) 62}.

\bibitem{Barrow2007}
J.D.~Barrow, R.~Maartens and C.G.~Tsagas, \emph{Cosmology with inhomogeneous
  magnetic fields}, {\emph{Physics Reports} {\bfseries 449} (2007) 131}.

\bibitem{Axel2021}
A.~Brandenburg, Y.~He, T.~Kahniashvili, M.~Rheinhardt and J.~Schober,
  \emph{Relic gravitational waves from the chiral magnetic effect},
  {\emph{Astrophys. J.} {\bfseries 911} (2021) 110}.

\bibitem{Kunze2013}
K.E.~Kunze, \emph{Cosmological magnetic fields}, {\emph{Plasma Phys. Control.
  Fusion} {\bfseries 55} (2013) 124026}.

\bibitem{Starobinsky1980}
A.A.~Starobinsky, \emph{A new type of isotropic cosmological models without
  singularity}, {\emph{Phys. Lett. B} {\bfseries 91} (1980) 99}.

\bibitem{Guth1981}
A.H.~Guth, \emph{Inflationary universe: A possible solution to the horizon and
  flatness problems}, {\emph{Phys. Rev. D} {\bfseries 23} (1981) 347}.

\bibitem{Linde1982}
A.~Linde, \emph{A new inflationary universe scenario: A possible solution of
  the horizon, flatness, homogeneity, isotropy and primordial monopole
  problems}, {\emph{Physics Letters B} {\bfseries 108} (1982) 389}.

\bibitem{Albrecht1982}
A.~Albrecht and P.J.~Steinhardt, \emph{Cosmology for grand unified theories
  with radiatively induced symmetry breaking}, {\emph{Phys. Rev. Lett.}
  {\bfseries 48} (1982) 1220}.

\bibitem{Garretson1992}
W.D.~Garretson, G.B.~Field and S.M.~Carrol, \emph{Primordial magnetic fields
  from pseudo goldstone bosons}, {\emph{Phys. Rev. D} {\bfseries 46} (1992)
  5346}.

\bibitem{Ratra1992}
B.~Ratra, \emph{Cosmological "seed" magnetic field from inflation}, {\emph{Ap.
  J. Lett.} {\bfseries 391} (1992) L1}.

\bibitem{Dolgov1993}
A.D.~Dolgov, \emph{Breaking of conformal invariance and electromagnetic field
  generation in the universe}, {\emph{Phys. Rev. D} {\bfseries 48} (1993)
  2499}.

\bibitem{Bertolami1999}
O.~Bertolami and D.F.~Mota, \emph{Primordial magnetic fields via spontaneous
  breaking of lorentz invariance}, {\emph{Phys. Lett. B} {\bfseries 455} (1999)
  96}.

\bibitem{Bertolami2005}
O.~Bertolami and R.~Monteiro, \emph{Varying electromagnetic coupling and
  primordial magnetic fields}, {\emph{Phys. Rev. D} {\bfseries 71} (2005)
  123525}.

\bibitem{Bamba2020}
K.~Bamba, E.~Elizalde, S.D.~Odintsov and T.~Paul, \emph{{Inflationary
  magnetogenesis with reheating phase from higher curvature coupling}},
  {\emph{JCAP} {\bfseries 04} (2021) 009}
  [\href{https://arxiv.org/abs/2012.12742}{{\ttfamily 2012.12742}}].

\bibitem{Chakraborty2022}
S.~Chakraborty, S.~Pal and S.~SenGupta, \emph{Hilltop inflation and generation
  of helical magnetic field}, {\emph{Universe} {\bfseries 8} (2022) 26}.

\bibitem{Saketh2020}
R.~Kothari, M.V.S.~Saketh and P.~Jain, \emph{Torsion driven inflationary
  magnetogenesis}, {\emph{Phys. Rev. D} {\bfseries 102} (2020) 024008}.

\bibitem{Gasperini1995}
M.~Gasperini, M.~Giovannini and G.~Veneziano, \emph{Primordial magnetic fields
  from string cosmology}, {\emph{Phys. Rev. Lett.} {\bfseries 75} (1995) 3796}.

\bibitem{Martin2008}
J.~Martin and J.~Yokoyama, \emph{Generation of large-scale magnetic fields in
  single-field inflation}, {\emph{JCAP} {\bfseries 01} (2008) 025}.

\bibitem{Atmjeet2014}
K.~Atmjeet, I.~Pahwa, T.R.~Seshadri and K.~Subramanian, \emph{Cosmological
  magnetogenesis from extra-dimensional gauss bonnet gravity}, {\emph{Phys.
  Rev. D} {\bfseries 89} (2014) 063002}.

\bibitem{Capoziello2011}
S.~Capozziello and M.~De~Laurentis, \emph{Extended theories of gravity},
  {\emph{Phys. Rep.} {\bfseries 509} (2011) 167}.

\bibitem{Clifton2012}
T.~Clifton, P.G.~Ferreira, A.~Padilla and C.~Skordis, \emph{Modified gravity
  and cosmology}, {\emph{Phys. Rep.} {\bfseries 513} (2012) 1}.

\bibitem{Bertol2011}
O.~Bertolami, \emph{What if ... general relativity is not the theory?},
  {\emph{Mem. S. A. It.} {\bfseries 75} (2011) 282}
  [\href{https://arxiv.org/abs/1112.2048}{{\ttfamily 1112.2048}}].

\bibitem{Bertolami2007}
O.~Bertolami, C.G.~Boehmer, T.~Harko and F.S.N.~Lobo, \emph{Extra force in f(r)
  modified theories of gravity}, {\emph{Phys. Rev. D} {\bfseries 75} (2007)
  104016}.

\bibitem{Gomes2017}
C.~Gomes, J.G.~Rosa and O.~Bertolami, \emph{Inflation in non-minimal
  matter-curvature coupling theories}, {\emph{JCAP} {\bfseries 06} (2017) 021}.

\bibitem{Alho2011}
A.~Alho and F.C.~Mena, \emph{Pre-inflationary homogeneization of scalar field
  cosmologies}, {\emph{Phys. Lett. B} {\bfseries 703} (2011) 537}.

\bibitem{Bertolami2014}
O.~Bertolami and J.~P\'aramos, \emph{Modified friedmann equation from
  nonminimally coupled theories of gravity}, {\emph{Phys. Rev. D} {\bfseries
  89} (2014) 044012}.

\end{thebibliography}\endgroup

\end{document}